# High-Brightness Photocathode Electron Sources

Christian Travier
Laboratoire de l'Accélérateur Linéaire
IN2P3-CNRS et Université de Paris-Sud
Bâtiment 200
F-91405 Orsay

## Abstract

Most present and future electron accelerators require bright sources. Invented less than ten years ago, the photo-injector the principle of which is briefly recalled, has already demonstrated that it can provide very bright beams. In this paper, the most advanced photo-injector projects are reviewed, their specific features are outlined, and their major issues are examined. The state-of-the-art in photocathode and laser technologies is presented. Beam dynamics issues are also considered since they are essential in the production of bright beams. Finally, the question of the maturity of photo-injector technology is addressed.

# INTRODUCTION

Numerous applications of electron linacs require high-brightness sources. These include high-energy linear colliders[1], short wavelength free electron lasers[2], wakefield accelerator experiments[3], new accelerator scheme test facilities[4], drive beam for two-beam accelerators (TBA)[5], coherent radiation sources[6], radiochemistry[7], ...

The brightness being proportional to the peak current divided by the square of the normalized emittance, bright electron sources require intense beams (high charge and short pulse) and small emittances. Table 1 indicates qualitatively the needs for the different applications, since they do not all require the same kind of beams. Figure 1 shows the normalized brightness needed by linear colliders and FEL applications. It also displays the present state-of-the-art of conventional injectors (DC gun + bunchers), thermionic RF guns and photo-injectors. This plot tends to show that a photo-injector allows on average two (respectively one) orders of magnitude improvement in brightness when compared to a conventional injector (respectively thermionic RF gun).

After a brief description of photo-injector principle and a short historical overview, this paper focuses on the nine most advanced projects. General remarks concerning current trends of R&D in the field of photocathodes, lasers, and beam dynamics in a photo-injector are then discussed.

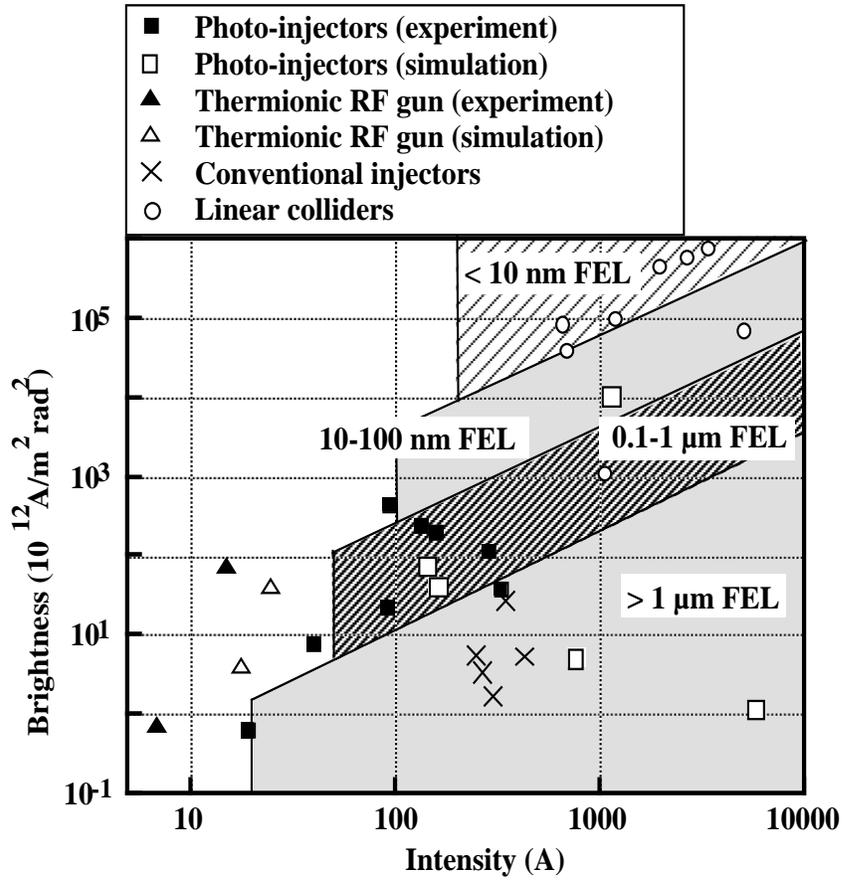

Figure 1: Brightness required for FEL and linear colliders

Table 1: Beam requirements for future applications

|  | High charge | Low emittance | Short pulse | Other |
|---|---|---|---|---|
| Linear colliders | * | *** | *** | polarized, asymmetric |
| FEL | ** | ** | * | high average power, very good phase and amplitude stability |
| TBA, RF sources | *** |  | ** |  |
| Wakefield accelerators | *** |  | ** |  |
| New acceleration schemes |  | ** | * |  |
| Coherent radiation sources |  | * | *** | high average power |
| Radiochemistry | *** |  | ** |  |

# PHOTO-INJECTOR PRINCIPLE

To increase the brightness of an electron source, it is necessary to increase its peak current while keeping a very small transverse emittance. This leads to the use of high electric fields to reduce the influence of space charge forces. Since DC voltages in a gun are limited to a few hundred kilovolts, it is more appropriate to use RF fields to extract high peak current from a cathode. Following this idea, G. Westenskow and J. Madey designed and operated in 1985, the first microwave gun consisting of a thermionic cathode located in an S-band RF cavity[19]. Meanwhile, the necessity of high peak current implies the use of short pulses. The electronic switching of the grid in a conventional DC gun does not allow one to produce pulses shorter than a few hundred picoseconds. To obtain shorter pulses, it is natural to think about optical switching. A short laser pulse illuminating a photocathode provides an almost ideal way to produce such short pulses. It was first experimented with at Los Alamos by Fraser and Sheffield[8]. The combination of acceleration in a high RF field and generation of electrons by short laser pulses hitting a photocathode make a quasi perfect bright injector. Today, lasers are able to produce very short pulses (down to less then 1 ps), photocathodes can deliver high current densities (several thousands of kA/cm$^2$) and RF cavities can sustain electric fields as high as 100 MV/m, so that RF photo-injectors can reach very high brightness.

This potentiality of photo-injectors to produce bright beams has therefore boosted their development. Since their invention 10 years ago, the number of photo-injectors has rapidly increased and currently exceeds 30. Table 2 gives a summary of the main breakthroughs occured during the last ten years in this quickly progressing field of RF photo-injectors.

Table 2: Main breakthroughs in photo-injector history

| Date | Event | Reference |
|---|---|---|
| 1985 | First photo-injector at Los Alamos | 8 |
| 1988 | First FEL driven by photo-injector at Stanford | 9 |
| 1988 | Emittance compensation theory by B. Carlsten | 10 |
| 1989 | First S-band photo-injector at Brookhaven | 11 |
| 1989 | Analytic theory of photo-injector beam dynamics by K.J. Kim | 12 |
| 1990 | First 144 MHz photo-injector at CEA | 13 |
| 1992 | First 433 MHz high duty cycle photo-injector at BOEING | 14 |
| 1992 | First photocathode in a superconducting RF cavity at Wuppertal | 15 |
| 1993 | UV FEL driven by photo-injector at Los Alamos | 16 |
| 1993 | First sub-picosecond laser driven photo-injector at LAL | 17 |
| 1994 | New analytic theory of photo-injector beam dynamics by L. Serafini | 18 |

For the sake of completness, it should be noted here, that it is also possible to use a photocathode in a DC gun, thus making a DC photo-injector. Such a device could in principle deliver short pulses. However, due to the moderate accelerating gradient, only very low charge can be extracted in short pulse. Practically, DC photo-injectors are used to produce intense pulses of a few hundred picoseconds to a few nanoseconds, and therefore still necessitate RF bunchers. The DC photo-injector has been extensively studied at SLAC, where a 120 kV gun using a GaAs photocathode is delivering daily polarized electrons to the SLC[20].

# REVIEW OF ADVANCED PHOTO-INJECTORS

The most advanced photo-injector projects are reviewed below. The main features of each project are discussed and table 3 summarizes their parameters. Except for ANL and MIT, these results correspond to experimental data and represent consistent sets of typical parameters. One should therefore be very careful while comparing these data since some of them (eg. emittance) are rather difficult to measure.

### Los Alamos National Laboratory

As already mentioned, the first photo-injector was designed and built at Los Alamos by J. Fraser and R. Sheffield[8]. After this first prototype, LANL has built APEX[21] and AFEL[22] photo-injectors. These devices are sophisticated guns made of several RF cells and using the emittance compensation scheme devised by B. Carlsten[10]. The APEX gun brightness was so high that it allowed the first UV FEL lasing on a linac[16]. The AFEL gun is made of 11 cells (see figure 2) and produces a very bright electron beam used to drive a very compact FEL. Present studies include a detailed understanding of emittance measurements for these very bright beams.

### Brookhaven National Laboratory

The BNL gun design[23] shown in figure 3 is the most popular one, since it has already been reproduced 10 times. The careful design of the cavity shape is intended to completely suppress higher spatial harmonics of the field, thus minimizing the non-linear emittance. The high gradient operation (up to 100 MV/m) allows one to produce the very small emittance beams needed by the FEL and advanced accelerator physics experiments done at the Brookhaven ATF facility.

The most oustanding recent result is the convenient use of a magnesium cathode, that proved both to have a relatively good quantum efficiency ($5 \times 10^{-4}$) and be very robust (lifetime over 5000 hours)[24].

This one and a half cell gun is now being replaced by a three and a half cell gun jointly designed and fabricated by BNL and Grumman[25] and that can work at very high duty cycle (1%). A new laser system is also being assembled (see the parameters in table 5). Together with SLAC, UCLA, NRL, and LANL, BNL is now designing an inexpensive gun that would allow smaller laboratories or universities and smaller groups inside big laboratories to afford such a bright gun for any type of application.

Table 3: Parameters for the main photo-injectors

| Parameter | CEA | LANL | ANL | BNL | CERN | KEK | UCLA | LAL | MIT |
|---|---|---|---|---|---|---|---|---|---|
| Purpose | FEL | FEL | Wakefield accelerator | Advanced accelerator | Linear collider | Linear collider | FEL | Linear collider | High gradient |
| First operation | 1990 | 1992 | - | 1989 | 1990 | ? | ? | 1993 | - |
| Running time (h) | 3200 | 1000 | 0 | 5000 | 1800 | ? | ? | 100 | 0 |
| Number of cavities | 1 | 11 | 1 | 1.5 | 1.5 | 1 | 1.5 | 2 | 1.5 |
| Frequency (MHz) | 144 | 1300 | 1300 | 2856 | 2998 | 2856 | 2856 | 2998 | 17136 |
| Macropulse ($\mu$s) | 20-160 | 20 | 8 | 3 | 2.5 | 2 | 4 | 3 | 0.05-1 |
| Repetition rate (Hz) | 1-10 | 20 | 30 | 1.5-6 | 10 | 5 | 10 | 12.5 | 1-4 |
| Vacuum RF off (nTorr) | 1 | 0.2 | 100 | 10 | 0.1 | ? | ? | 10 | 2 |
| Vacuum RF on (nTorr) | 2.5 | 1 | ? | 50 | 1 | ? | ? | 50 | 10 |
| Cathode field (MV/m) | 28 | 20 | 90 | 70 | 100 | 40 | 83 | 50 | 250 |
| Cathode | $Cs_3K_2Sb$ | $Cs_2Te$ | Y | Cu | $Cs_2Te$ | CsSb | Cu | Cu | Cu |
| Quantum efficiency (%) | 3 | 5 | 0.05 | 0.001 | 2 | ? | ? | $5\times10^{-4}$ | 0.001 |
| Lifetime | 1 h | months | ? | $\infty$ | 70 h | ? | $\infty$ | $\infty$ | ? |
| Laser | YAG | YLF | Kr-F | YAG | YLF | YAG | YAG | Ti:sa | Ti:sa |
| Wavelength (nm) | 532 | 263 | 248 | 266 | 262 | 532 | 266 | 260 | 260 |
| Pulse length FWHM (ps) | 20-50 | 6 | 3 | 15 | 8 | 10 | 4 | 0.5 | 2 |
| Energy single pulse ($\mu$J) | 20 | 50 | 12000 | 300 | 1 | 100 | 300 | 250 | 200 |
| Spot size FWHM (mm) | 2-7 | 4 | 20 | 0.1-1 | 5 | ? | 0.6 | 4 | 1 |
| Energy (MeV) | 2 | 16 | 1.7 | 3 | 4 | 0.9 | 3.5 | 2.2 | 2.8 |
| Charge (nC) | 0.5-5 | 3 | 100 | 0.5 | 4 | 3.2 | 0.5 | 0.11 | 0.1-1 |
| Pulse length FWHM (ps) | 20-50 | 20 | 14 | 11 | 13 | ? | 5 | ? | 1.5 |
| Jitter (ps) | 3 | $<1$ | $<10$ | $<1$ | $<1$ | ? | ? | ? | 2 |
| Normalized rms emittance ($\pi$ mm mrad) | 4@1nC | 5 | 130 | 4 | ? | ? | 10 | ? | 3@1nC |

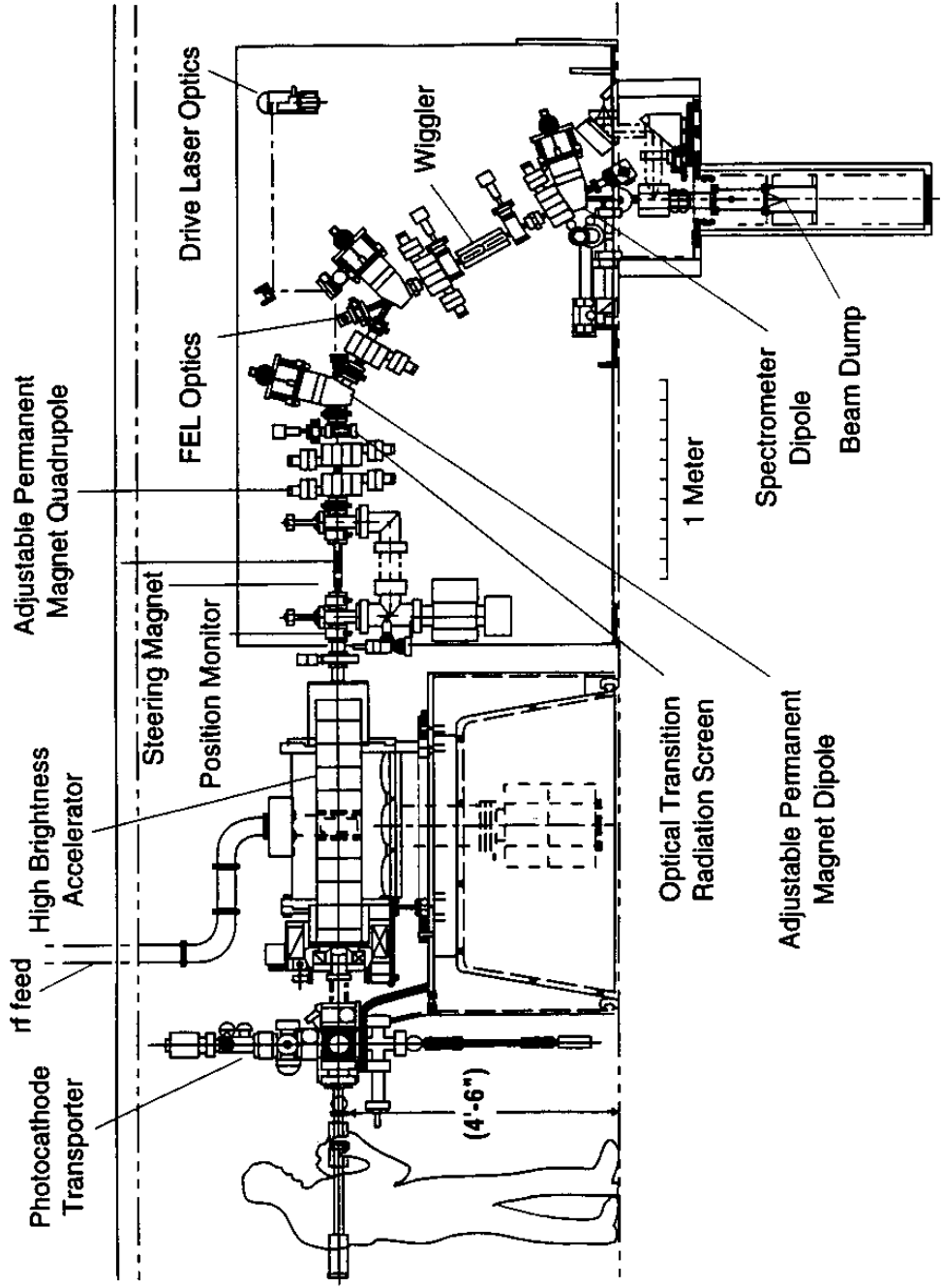

Figure 2: Schematic of the AFEL experiment

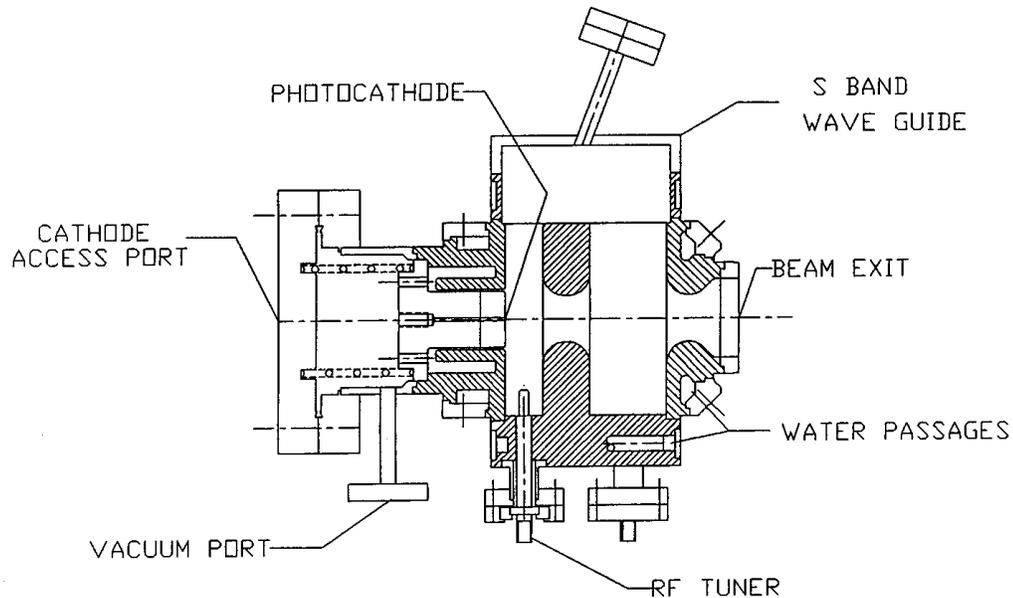

Figure 3: BNL one and a half cell gun

## CEA at Bruyères-le-Châtel

At CEA, the photo-injector is made of one 144 MHz cavity and is used as the electron source for the infra-red high power FEL[13]. Recent emphasis has been put on improving the stability and reliability of the system. An amplitude feedback system is being developed to improve the laser stability.

## CERN

The CTF (CLIC Test Facility) was built at CERN to test some components of the CERN Linear Collider project[26] based on the concept of the two-beam accelerator. In order to generate the RF power at 30 GHz, necessary to obtain the high accelerating gradient needed for the 30 GHz CLIC accelerating section, a train of short intense electron bunches is accelerated, and then produces RF power through electromagnetic interaction with a so-called tranfer structure. This train of intense short electron pulses is generated by a BNL type RF gun, using a $Cs_2Te$ photocathode. Recently a train of 24 pulses, 14 ps long and 6.2 nC each (at gun exit) led to the production of 40 MW of 30 GHz RF power[27,28]. Figure 4 shows a schematic of the experiment.

When running with a single pulse, a charge as high as 26 nC was extracted from this photo-injector. The $Cs_2Te$ photocathode presents a good combination of a very high quantum efficiency (2-5 %) and a long lifetime (several months).

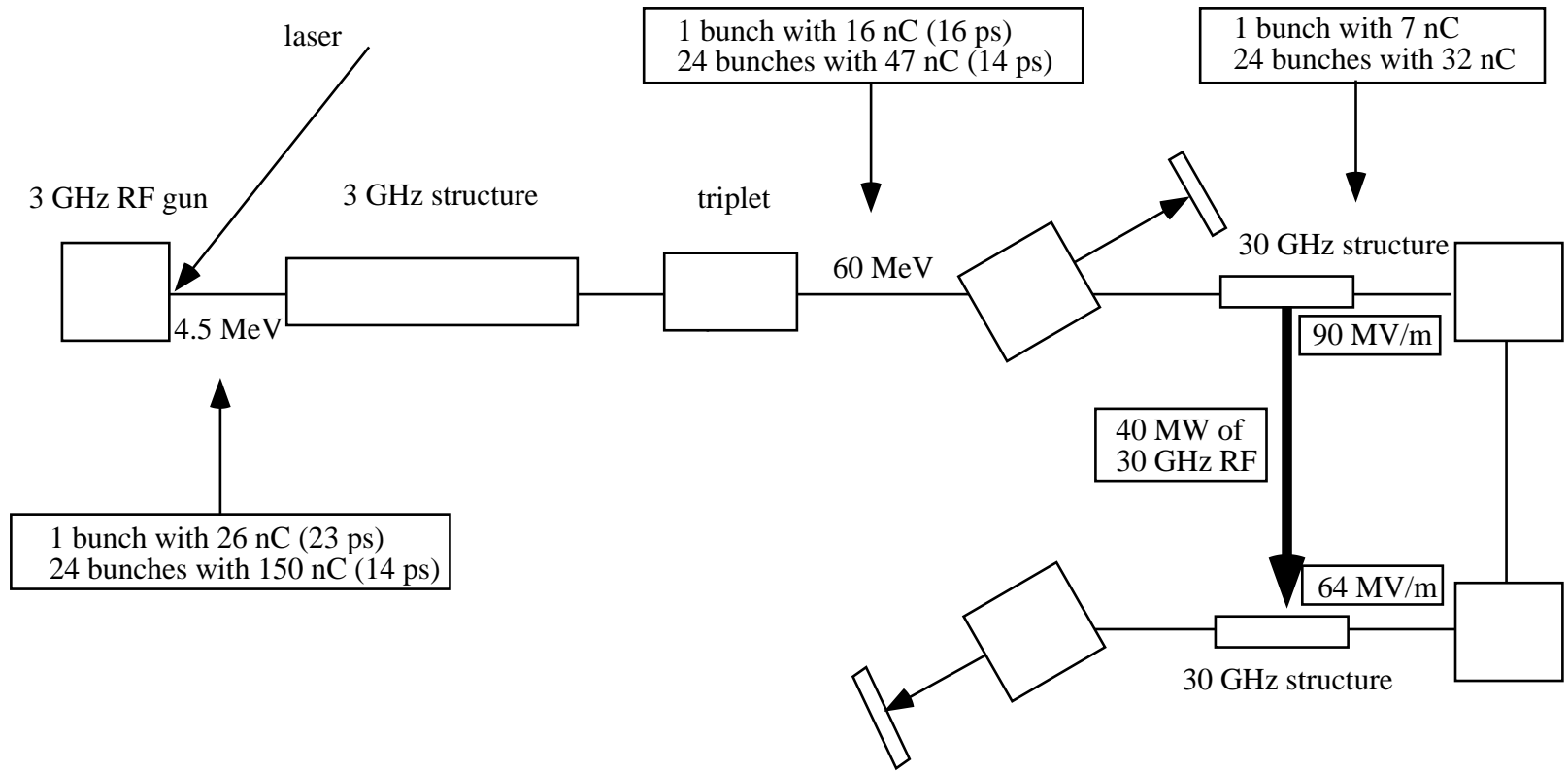

Figure 4: Schematic layout of CLIC Test Facility

## KEK

A 1 cell S-band RF gun was developped at KEK in view of JLC linear collider[29]. This gun is the only one working with relatively high accelerating gradient ($> 40$ MV/m) and using an alcaline cathode (CsSb). A special emphasis has therefore been put on the vacuum system design with the use of NEG pumping. Since linear colliders need polarized sources, KEK is working on the subject of polarized photocathodes[30] and polarized RF gun in collaboration with SLAC[38].

## University of California at Los Angeles

The UCLA RF gun is a modified BNL gun. It allows 70 degree laser illumination of the cathode which produces an enhancement of the quantum efficiency when compared to normal incidence. Extensive measurements and comparison to simulation were made and proved to be satisfactory[31].

## Laboratoire de l'Accélérateur Linéaire Orsay

A two decoupled cell S-band RF gun was recently put into operation at LAL[17]. The cathode is illuminated by a Ti:sapphire laser shown in figure 5, that produces 0.5 ps pulses[32]. This project is the first sub-picosecond laser driven photo-injector. Though these preliminary results were obtained with a copper cathode, a dispenser cathode (WCaOBaO) used as a photoemitter will soon be installed in the gun. This cathode has a measured quantum efficiency of $5 \times 10^{-4}$ with UV light and at room temperature[33].

## Argonne National Laboratory

In order to do wakefield accelerator experiments, it is necessary to generate very intense and short electron bunches[3]. At ANL, a photo-injector designed to produce 100 nC pulses is being commissioned. To maintain a pulse as short as possible in spite of the enormous space charge forces, a concave shape of the laser wavefront is created[34]. To probe the wakefield excited by this intense pulse, a second pulse is generated by a 7 cell photo-injector[35]. An alternative design for this witness beam gun is a dielectric loaded cavity that allows one to produce a perfectly linear accelerating field[35].

## Massachussets Institute of Technology

The MIT photo-injector is a one and a half cell BNL type gun scaled at 17 GHz[36]. The use of such a high frequency makes very high accelerating gradient possible. 250 MV/m is envisaged for this gun presently under commissioning.

# PHOTOCATHODES

A good photocathode for photo-injector operation should ideally have a high quantum efficiency ($> 1$ %) at infra-red or visible wavelength, have a long lifetime ($>$ several months) under moderate vacuum conditions, and be easy to prepare and install in the gun cavity. Such a perfect cathode does not exist yet, but progress was made recently, especially with $Cs_2Te$ and Mg cathodes. Table 4 shows the most commonly used photocathodes and a summary of their main advantages and drawbacks.

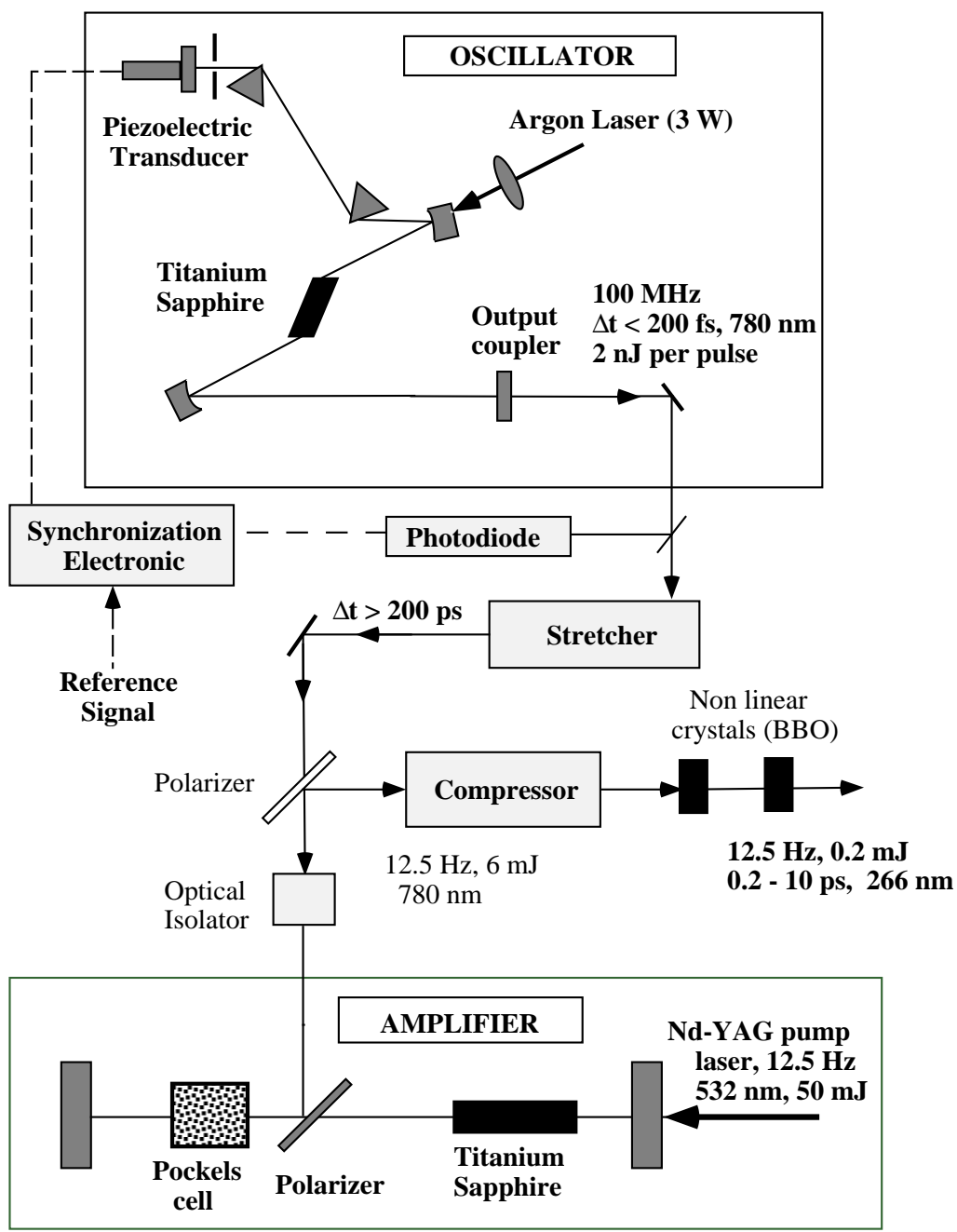

Figure 5: LAL Orsay Ti:sapphire femtosecond laser

Table 4: Main photocathodes used in photo-injectors

| Cathode | Advantage | Drawback |
|---|---|---|
| $Cs_3Sb$, $CsK_2Sb$ | high quantum efficiency<br>0.5 $\mu$m laser | difficult to prepare<br>short lifetime<br>do not sustain very high fields<br>expensive preparation chamber and transfer system<br>need good vacuum |
| $Cs_2Te$ | high quantum efficiency<br>long lifetime<br><br>sustains high fields | needs UV laser<br>expensive preparation chamber and transfer system<br>response to train? |
| Cu, Y, Mg | no preparation chamber<br><br>long lifetime<br>fast response<br>sustain very high fields<br>sustain bad vacuum | low quantum efficiency (except Mg)<br>need UV laser |
| $LaB_6$, WCaOBaO | no preparation chamber<br>long lifetime | low quantum efficiency<br>need UV laser<br>need to be heated prior to operation |

Specific issues are to be considered for two particular applications: polarized sources and superconducting photo-injectors.

- The only photocathode presently available to generate polarized electron is GaAs. Bulk GaAs gives polarization of 40 %, where strained layers of GaAs can lead to polarization as high as 90 %. This type of cathode under extensive study[37], has never been used in an RF gun so far[38]. Many limitations might appear in doing so in order to produce short intense polarized electron beams. Besides technological problems such as sensitivity of the cathode to vacuum conditions or high gradient, there are more fundamental problems such as the charge limit effect discovered and explained at SLAC[20], and the time response to picosecond pulses[39]. These problems are now being investigated since polarized electron sources is an important challenge of any linear collider project.

- Using superconducting cavities for an RF gun is interesting in some cases, as for example very high duty cycle machines. In this case, one has to find a photocathode that can work in a superconducting environment, i.e. introducing the following specific requirements: work function below that of niobium (4.6 eV) to avoid photoemission from the surface of the cavity, and very low e.m. losses at cryogenic temperature in order to preserve the very high quality factor of the cavity. The first experimental work done at Wuppertal has shown that a $CsK_2Sb$ cathode has better quantum efficiency at cryogenic temperature than at room temperature. It also showed that if the cathode is too thick, it is not superconducting and thus leads to very high ohmic losses[15]. Following this work, studies are now made at Milan to produce very thin (< 30 nm) cathodes, that can become superconducting by proximity effects[40].

The choice of the photocathode to be used depends on the type of applications. It depends of course on the charge required from a single pulse, but also on the pulse format via the existence or not of a suitable laser. The typical cases are the following:

- if a single pulse of charge below 5 nC, is required, a Mg photocathode is probably the best choice.

- if train a pulses of charge less than a few nC are required, then one should probably go to $Cs_2Te$.

- for very high repetition rate or very high duty cycle and high charge, it is probably difficult to avoid $CsK_2Sb$.

# LASERS

One of the key components of a photo-injector system is the laser. Amplitude, phase and position stability of the electron beam depend almost completely on the laser performances. A laser is typically made of an oscillator that generates a continuous train of pulses of low energy (few nJ). This oscillator is synchronized via an appropriate electronic system to a sub-harmonic of the RF frequency (typically 100 MHz). One single pulse or a train of a few hundred pulses is then selected through a Pockells cell and amplified. There exist several types of amplifiers (single pass, multiple pass, regenerative, ...) and according to the energy desired, it might be necesssary to have several amplifier stages. When the oscillator pulse is very short, one has to extend the pulse temporally before amplification to avoid damage of the amplifier cavity components. The pulse is then compressed back to its original duration. This technique is called chirped pulse amplification. The oscillators used so far for photo-injector applications produce infra-red light. To obtain usable light for the photo-cathode, it is therefore necessary to generate higher harmonics. This is done by using non-linear crystals, with a typical efficiency of 10-15% from the fundamental to the third or fourth harmonic.

Nd:YAG and Nd:YLF (eg. reference 41) are the most commonly used systems in existing photo-injectors. They typically produce pulses of 6-15 ps, with up to 300 $\mu$J of energy in a single pulse. More recently the advent of Ti:sapphire has opened a way down towards the very short pulses (eg. reference 32). Table 5 gives the performances of the present BNL Nd:YAG laser together with the expected performances of the new Nd:YAG laser now under construction.

Table 5: Parameters of the present and future BNL Nd:YAG lasers

|  | Present | Future |
|---|---|---|
| **Single-Pulse Mode** | | |
| Energy: | | |
| IR (total) | 20 mJ | ≥10 mJ |
| IR into $2\omega$ crystal | 8 mJ | ≤ 3 mJ |
| Green | 1.6 mJ | ≤1 mJ |
| UV | 100 $\mu$J | ≤300 $\mu$J |
| Repetition rate | 1.5, 3 Hz | 1.5, 3 Hz |
| Pulse duration (FWHM): | | |
| Oscillator IR | 100 ps | 17 ps |
| Amplified IR | 17-25 ps | 24 ps |
| Green | 12-18 ps | 19 ps |
| UV | 9-13 ps | 15 ps |
| Beam diameter (FWHM) on cathode | 0.05-3 mm | 0.1, 0.5, 1 mm |
| Shot to shot stability (on cathode, rms): | | |
| Timing | ≥ 3 ps | 1 ps |
| Energy | 10% | 2% |
| Pointing (1 mm beam diameter) | ≥ 100 $\mu$m | 20 $\mu$m |
| Pointing (0.1 mm beam diameter) | ≥ 40 $\mu$m | 20 $\mu$m |
| **Pulse-Train Mode** | | |
| Pulses in amplified train | 1-100 | 1-200 (100) |
| Pulse spacing | 24.5 ns | 12.25 (24.5) ns |
| IR energy per pulse | 1 mJ | 0.75 (1.5) mJ |
| Pulse energy variation across train (rms) | 8% | 2% |
| UV energy per pulse on cathode |  | 40 (80) $\mu$J |

The recent progress of diode pumping made possible the design of very compact, stable, reliable and relatively cheap Nd:YLF oscillators. More work has yet to be done to improve the performances of amplifiers and harmonic generation in terms of output energy, amplitude stability and beam quality.

Sophisticated feedback and feedforward loops are now being envisaged to improve the different types of stability. Temporal and spatial filters allow, in principle, the production of any longitudinal and transverse profile, at the expense of energy. These features are interesting for experimental tests of theoritical schemes developed to reduce the transverse emittance.

The R&D on short pulse lasers relevant for photo-injector application is described in detail in references 42 and 43.

## BEAM DYNAMICS

RF gun beam dynamics was worked out in an analytical manner by K.J. Kim[12]. This simple model provides the gun designer with handy formulae for the different gun parameters, and are especially useful to understand the scaling of these parameters with such variables as the bunch length, the spot size on cathode, the peak accelerating field or the bunch charge. An improved model giving more accurate results, especially for the transverse bunch dimension was recently derived by L. Serafini[18].

Most of the theoritical work done on RF gun beam dynamics concerns the possibility of obtaining smaller emittances, either by compensating the correlated emittance generated mainly in the first cell, or by removing the causes of extra emittance growth. Most of these techniques are reviewed in reference 44. Not all of them were experimentally proven or even tested. The most successfull one is the compensation of the correlated linear space charge induced emittance by the use of a magnetic focusing solenoid, due to B. Carlsten[10]. This technique experimentally proven at Los Alamos, allows a ten fold reduction of the emittance. Once thought unapplicable at high frequencies and high gradient, it will soon be implemented at BNL[45].

Besides emittance compensation schemes, recent ideas being studied include travelling wave RF guns[46] and asymmetric guns for linear colliders[47].

## CONCLUSION

Ten years after its invention, the photo-injector has reached the point where it is used daily at user facilities (e.g. BNL), and where it opened the way to new results such as UV FEL and the generation of 30 GHz power. However, can it already be considered as a reliable technology?

Before answering this question, one can try to classify what are the main problems and difficulties encountered by the people who are operating photo-injectors. Table 6 shows the result of a survey the author made through the people in charge of the different photo-injectors that were presented above. It definitely shows that the laser is the most important issue. This is rather normal, since the lasers needed by photo-injector applications are quite specific and therefore were not ready in the catalogs waiting to be used. They need to undergo specific developments, and this of course takes time, effort and money.

Table 6: Survey on photo-injector current issues

| | |
|---|---|
| Laser stability (amplitude, phase, ...) | 38% |
| Photocathode (QE, lifetime, cost, RF contact, ...) | 15% |
| High gradient (dark current, ...) | 13% |
| Miscellaneous | 11% |
| Vacuum | 8% |
| Synchronization RF/laser | 5% |
| Cost | 5% |
| Instrumentation | 5% |

One of the questions of the above mentioned survey was: "*Do you think photo-injector technology is mature enough to be used in a user facility for any kind of application? If not, when might it be?* In spite of the issues currently faced by photo-injector builders and operators, the general response to this question was "yes". The details of the different answers are summarized in table 7. Therefore, it seems that, provided that the remaining problems mainly concerning the laser stability and reliability are solved, the photo-injector is ready to replace the conventional DC gun + buncher injector for any kind of application.

Table 7: Answers to the question of maturity of the photo-injector technology

| |
|---|
| Yes |
| Yes, with a strong team of laser specialists |
| Yes, if enough money and manpower |
| Not perfect, but already used at some user facilities |
| Hope it will be soon |
| If compared to thermionic gun, no. OK in 96-97 |

# ACKNOWLEDGEMENTS

The data and information for the different projects were kindly provided by I. Ben Zvi, S.C. Chen, J.P. Delahaye, A. Fischer, S. Joly, I. Madsen, R. Sheffield, P. Schoessow, M. Yoshioka. Discussions with P. Georges were appreciated when writing the laser section. This paper benefited from the discussions of the source working group during the Workshop. I would also like to thank P. Schoessow and A. Fry who kindly helped me to print my transparencies during the last two days before my talk, and T. Garvey who carefully red this manuscript.